\documentclass[letterpaper, 10 pt, conference]{ieeeconf}
\IEEEoverridecommandlockouts
\usepackage{amsmath}
\usepackage{amssymb}
\usepackage{graphicx}
\usepackage{url}
\usepackage{xcolor}
\usepackage{booktabs}
\usepackage{array}
\usepackage{siunitx}

\newcolumntype{L}[1]{>{\raggedright\let\newline\\\arraybackslash\hspace{0pt}}m{#1}}
\newcolumntype{C}[1]{>{\centering\let\newline\\\arraybackslash\hspace{0pt}}m{#1}}
\newcolumntype{R}[1]{>{\raggedleft\let\newline\\\arraybackslash\hspace{0pt}}m{#1}}

\newcommand{\ra}[1]{\renewcommand{\arraystretch}{#1}}

\title{\LARGE \bf
Learning to predict metal deformations in hot-rolling processes
}

\author{R. Omar Chavez-Garcia$^{1}$, Emian Furger$^{2}$, Samuele Kronauer$^{2}$, Christian Brianza$^{2}$, \\ Marco Scarfò$^{3}$, Luca Diviani$^{2}$ and Alessandro Giusti$^{1}$
\thanks{This work was supported by INNOSUISSE under the project application number 33968.1 IP-ENG, and by the Swiss National Centre of Competence in Research Robotics.}
\thanks{$^{1}$R. Omar Chavez-Garcia and Alessandro Giusti are with the Dalle Molle Institute for Artificial Intelligence (IDSIA), USI-SUPSI, Lugano, Switzerland.
        {\tt\small omar@idsia.ch}}%
\thanks{$^{2}$Emian Furger, Samuele Kronauer, Christian Brianza and Luca Diviani are with the Institute for Mechanical Engineering and Materials Technology (MEMTI), DTI-SUPSI, Switzerland.}
\thanks{$^{3}$Marco Scarfò is with Montanstahl AG, Stabio, Switzerland.}
\thanks{Dataset, video demonstrations and additional information are available online at {\protect\url{ https://idsia-robotics.github.io/hot-rolling-deformation}}.}
}

\begin{document}

\maketitle

\begin{abstract}

Hot-rolling is a metal forming process that produces a workpiece with a desired target cross-section from an input workpiece through a sequence of plastic deformations; each deformation is generated by a stand composed of opposing rolls with a specific geometry. In current practice, the rolling sequence (i.e., the sequence of stands and the geometry of their rolls) needed to achieve a given final cross-section is designed by experts based on previous experience, and iteratively refined in a costly trial-and-error process. Finite Element Method simulations are increasingly adopted to make this process more efficient and to test potential rolling sequences, achieving good accuracy at the cost of long simulation times, limiting the practical use of the approach. We propose a supervised learning approach to predict the deformation of a given workpiece by a set of rolls with a given geometry; the model is trained on a large dataset of procedurally-generated FEM simulations, which we publish as supplementary material. The resulting predictor is four orders of magnitude faster than simulations, and yields an average Jaccard Similarity Index of $0.972$ (against ground truth from simulations) and $0.925$ (against real-world measured deformations); we additionally report preliminary results on using the predictor for automatic planning of rolling sequences.

\end{abstract}

\section{INTRODUCTION}

\begin{figure}
    \centering
    \includegraphics[width=\linewidth]{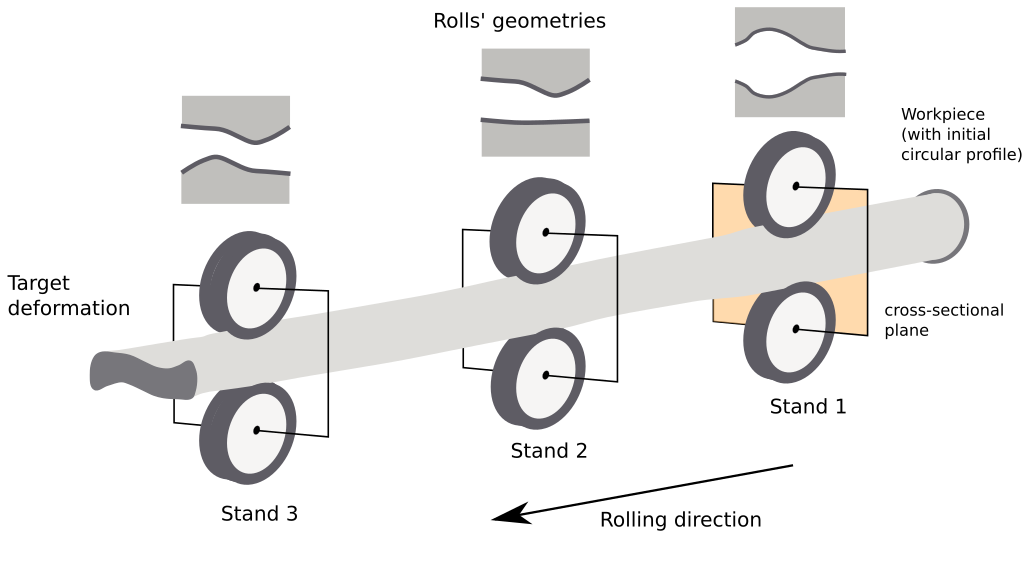}
    \includegraphics[width=0.74\linewidth]{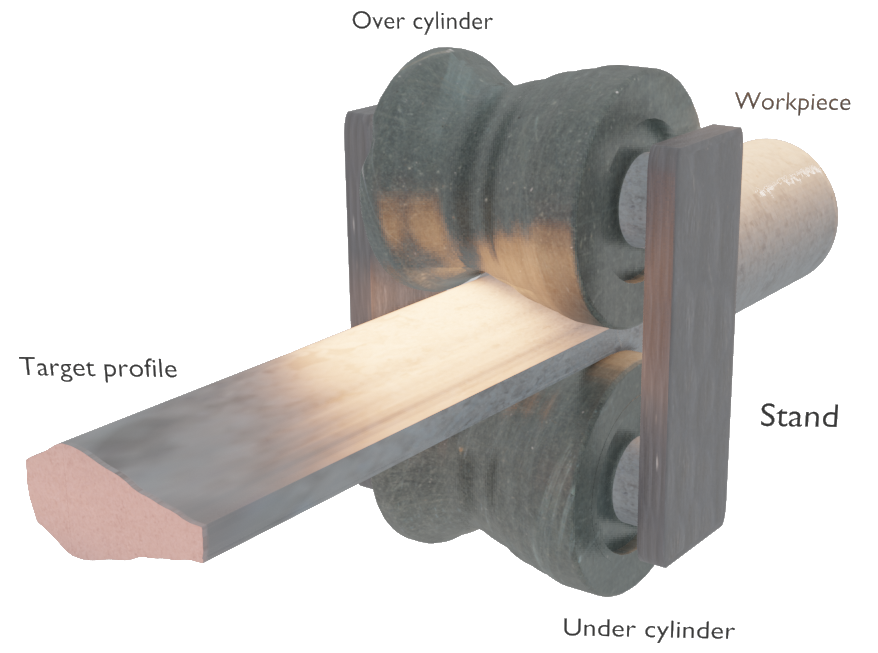}
    \includegraphics[width=0.98\linewidth]{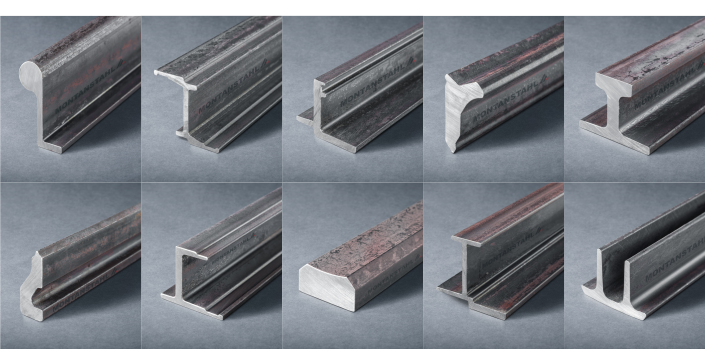}
    \caption{Top: Overview of a hot-rolling process: a workpiece with a circular cross-section is passed through a sequence of three stands, each equipped with two opposing rolls. 
    The final shape approximates a desired target. Middle: Close-up of a single-stand deformation (the first stand).
    Bottom: examples  of  real  profiles  generated  by  hot-rolling  deformations (courtesy of Montanstahl AG).
    }
    \label{fig:hot_rolling_pass}
\end{figure}

Hot-rolling is the process of reducing deforming the cross-sectional area of a workpiece at high temperatures, by applying compression forces delivered by rolls. The rolling sequence (see Figure~\ref{fig:hot_rolling_pass}) is the set of stands, and their respective roll geometries, that when applied in sequence on a metal rod, yield a final target profile. Hence, a rolling sequence comprises several single-stand configurations that progressively deform the workpiece according to their rolls' geometry.

Each deformation involves many complex physical phenomena, making it difficult to predict the reduction of the cross-sectional area of the workpiece due to the flow in the longitudinal direction and the amount of material flowing towards the side gaps (where rolls don't constrain the outlet shape).
As a consequence, designing a rolling sequence to obtain a given target profile is a difficult task. It involves choosing the correct number of stands and their configuration, including the geometry of rolls, ensuring that forces and material displacements in each stand do not exceed predetermined limits while yielding the desired profile.  Current industry practice relies on expertise from previously built profiles, followed by trial and error attempts with iterative adjustments, which result in high material and time costs. State-of-the-art processes rely on Finite Element Methods (FEM) to simulate complete or partial hot-rolling sequences~\cite{ataka2015}. Simulation results guide the designing process before production starts, which reduces the material costs and the setup time at the expense of long simulation times: FEM simulation of the deformation in a single stand requires several hours, especially in high material displacements on the side gaps.

We propose a data-driven approach to estimate, via Convolutional Neural Networks (CNN), the deformation of a workpiece by a set of rolls in a single-stand configuration. The trained model provides deformation estimates in a few milliseconds, and can be used in place of FEM simulation during rolling sequence design: this allows much faster iteration times and a more thorough exploration of the design space. 
We represent cross-section geometries as binary images. The model inputs comprise the shape of the input workpiece (inlet) and the shapes of the two opposing cylinders. The output is the cross-section of the resulting workpiece (outlet).  The model is trained with a large amount of procedurally-generated FEM simulations.

This approach constitutes the {\bf main contribution} of the paper, which is thoroughly validated against FEM simulations and real-world deformation data; results show that the estimator yields better accuracy than baselines captures and accurately models complex phenomena such as material flow.
As additional contributions: we publish a novel dataset of single-stand hot-rolling deformation FEM simulations used to train the model; we describe the procedural roll geometry generation strategy we used to produce it; we report preliminary experiments on the use of the predictor to automatically plan feasible multi-step rolling sequences that yield a given target cross-section.

After reviewing the related work in Section~\ref{sec:relatedwork}, we introduce the FEM simulation setup (Section~\ref{sec:hotrollingsim}) we used to generate the deformation dataset (Section~\ref{sec:datageneration}). Then we describe the proposed data-driven deformation estimator in Section~\ref{sec:deformationestimation} and report its evaluation against simulated (Section~\ref{sec:results}) and real-world (Section~\ref{sec:resultsreal}) data.
In Section~\ref{sec:planningaplication}, we show preliminary results on the use of the proposed estimator for planning rolling sequences. Section~\ref{sec:conclusions} concludes the paper and outlines future work.

\section{RELATED WORK}
\label{sec:relatedwork}

A typical production chain for a hot-rolling process consists of 5 stages:  design, tools manufacturing, setup, execution, and verification.  The design process establishes the parameters (material, geometries). Tools manufacturing produces the tools (mainly rolls) to generate the desired profile. Setup concerns the installation and tuning of the manufactured tools. Hot-rolling execution adjusts the needed parameters and executes the hot-rolling process. Verification computes metrics to establish the accuracy of the result.
These five stages are performed iteratively until reaching the desired accuracy, which might result in several iterations over the designing and tools manufacturing processes~\cite{panjkovic2017}. Reducing the number of iterations, or guiding the designing process decreases prototyping resources, energy consumption, and production time, which results in a shorter time-to-market.

Modern techniques for optimizing hot-rolling processes rely on ad hoc software for simulations, analysis, and optimization aiming at examining the interaction of the parts involved in the process before the production stage~\cite{ataka2015,liu20111}. These numerical tools are mainly based on the Finite Element Method and are widely used for process design and engineering of parts \cite{pittner2010}. Simulation provides a detailed analysis of the metal forming execution process and allows one to identify critical factors affecting it before executing such a process with real machinery.
The current simulation-based strategy involves the simulation of a sequence of single-stand configurations, which in turn simulates a whole rolling sequence.

Notably, a critical bottleneck in the simulation strategy is the time needed to simulate complex single-stand scenarios, e.g., simulations where large amounts of material are displaced. Consequently, as a hot-rolling process consists of a sequence of single-stand modules, long simulation times for such modules render complete hot-rolling sequences impracticable to simulate.
We can tackle this bottleneck through Machine Learning to perform deformation estimations of the long-time simulations, as has been done in other works to speed up time-consuming tasks such as image classification~\cite{krizhevsky2012}, segmentation~\cite{long2015}, and object detection~\cite{ren2015}. 

The prediction of physical interactions have been broadly studied, ranging from simple rigid object interplays to interactions at the particle level. These methods tackle the challenge of predicting how agents' actions affect the physical features of the environment.
Wu et al.~\cite{wu2017} propose a generative pipeline to learn rigid-object dynamics by interpreting and reconstructing the frames from a video stream using a convolutional inversion network. Finn et al.~\cite{finn2016} propose a pixel-level motion prediction using a Long Short-Term Memory network based on the conditioning of the agents' actions taken on a video stream. 
Both approaches tackle the problem of predicting the outcome of a complex physical process using Machine Learning.
Hot-rolling processes can also be represented as such a problem, where inlet and rolls represent the input, and the estimates represent deformation. In our context, we consider additional requirements that are not covered in the aforementioned approaches—notably, deformable bodies and physical models of the deforming material.
Additionally, developing differentiable physics engines to integrate into gradient-based learning methods to estimate position, speed, and mass, has shown promising results, albeit its limitations of only simulating rigid bodies~\cite{deavila2018}.

Mrowca et al.~\cite{mrowca2018} propose a novel hierarchical graph-based representation to model rigid and deformable bodies as a set of connected particles and a graph-convolutional neural network that learns physical predictions from this representation. Although this work shows promising results on predicting location, speed, and deformation of simple deformable bodies, its application on physical processes involving complex object interactions, such as hot-rolling deformation, is beyond their scope.

Several studies have been proposed to model physical processes involving the estimation of metal properties. Neural Networks based approaches can estimate melting point~\cite{guo2019}, bending strength \& hardness~\cite{koker2007}, and thermal conductivity ~\cite{lopez2015}. However, these methods only consider a few variables involved in the physical process or assume simplified scenarios. Kim \& Kim~\cite{kim2009} proposed using a shallow neural network in the hot-forging process to estimate the unfilled volume of a die cavity given the size of the workpiece. Yet, this method does not consider the importance that the geometry of the pieces has in this physical process.

Our data-driven strategy aims at estimating metal deformation generated by hot-rolling processes in FEM simulations. This estimator is modeled as a Convolutional Neural Network with residual blocks (an architectural unit that has shown to be very effective in many image-related problems)~\cite{he2016}.



\section{HOT-ROLLING SIMULATION}
\label{sec:hotrollingsim}

Generating data from the real hot-rolling process is unfeasible given the high costs in resources and the several hours employed to setup such process; simulating this process allows tackling, to some degree, these drawbacks. Another advantage of simulation is the ability to test a variety of hot-rolling scenarios, i.e., roll geometries, which would not be possible in the real hot-rolling process due to lack of production interest. To the best of our knowledge, there is no available dataset related to deformations in hot-rolling (or even cold rolling) metal forming processes.

A 3D FEM model was developed aimed at simulating a single-stand scenario. This model was validated using data from the real hot-rolling process. This validation showed that the cross-section of the output could be predicted with an average error of $4.67\%$~\footnote{The area error between simulation and the real piece is calculated according to $A_{\text{error}}=\frac{A_{\text{sim}}\bigcup A_{\text{real}}-A_{\text{sim}}\bigcap A_{\text{real}}}{A_{\text{real}}}$.}. 
Simulation time, even for a single-stand scenario with two vertical rolls, amounts up to \SI{25}{hours}~\footnote{3D simulation times were obtained from simulating a single-stand process (similar to the one in Figure~\ref{fig:hot_rolling_pass}:bottom) on a FEM simulation software for non-linear analysis (using an iterative sparse solver) on a single core Intel i7 machine.}.

However, given a large amount of required training data, it is necessary to decrease the computational time. A good compromise is represented by 2D simulations, which, compared to 3D simulations, can provide a large number of samples in a shorter time with a tolerable reduction of the quality of the results if only the cross-sectional area is considered. The development and assessment of the 2D model are detailed in Section~\ref{sec:fesimulationsdesc}.

\subsection{2D FEM Simulations}
\label{sec:fesimulationsdesc}

\begin{figure}
    \centering
    \includegraphics[width=\linewidth]{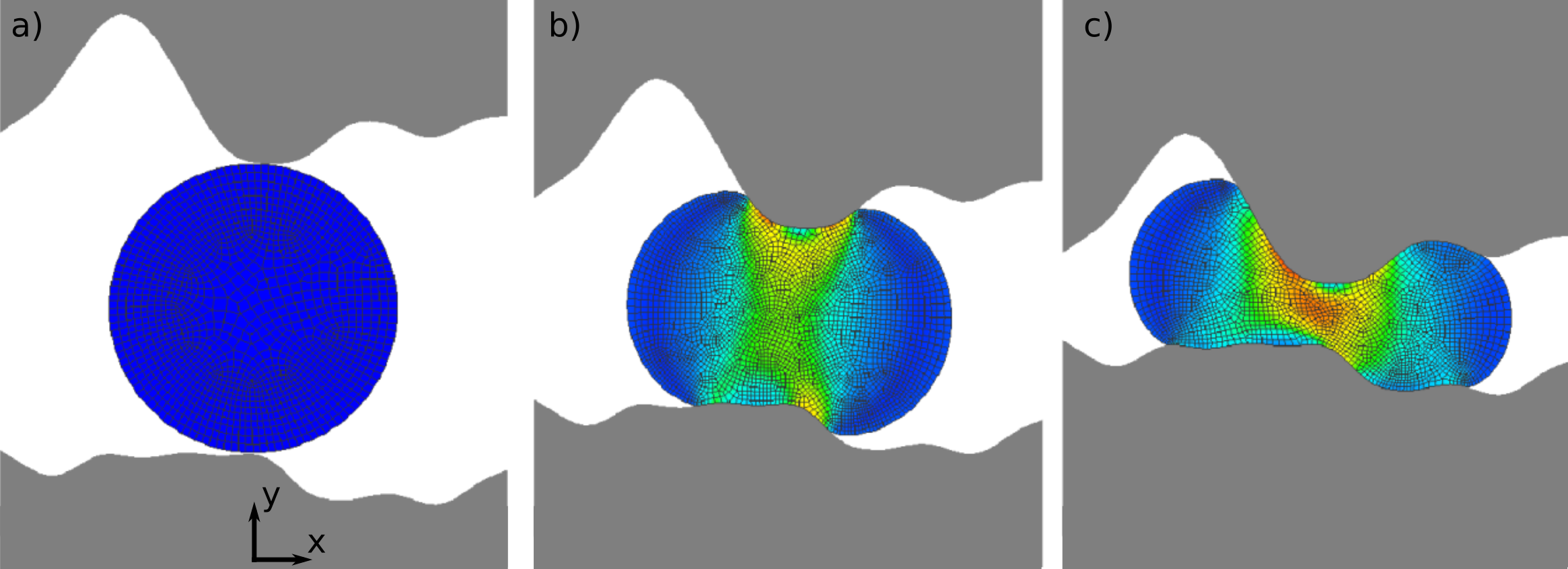}
    \caption{2D FEM model representing the hot-rolling process a) at the beginning of the simulation, b) when the rolls have traveled \SI{50}{\%} of the way, c) and at the end of the simulation.}
    \label{fig:screenshots2DFEM}
\end{figure}

The 2D FEM model of a single-stand scenario is composed of three bodies: 2 rolls and the workpiece (see Figure~\ref{fig:screenshots2DFEM} and takes place at the cross-sectional plane of the stand (see Figure~\ref{fig:hot_rolling_pass}:bottom). The workpiece at the beginning of the simulation has a circular cross-section, and the initial temperature is constant everywhere. The rolls are modeled as rigid bodies; therefore, no material model needs to be defined. The workpiece is modeled with 2D quadrilateral elements with a plane stress formulation. The used material model is defined by Shida~\cite{shida1969}, which describes the behavior of the material as a function of the chemical content of steel (in this case, S235JR) at a given temperature and strain rate range.
The supplier provides the chemical content to each batch, and an average of over $24$ samples was considered to determine an accurate chemical content. Friction is modeled with the law of Coulomb ($\mu=0.6$) between the workpiece and the two rolls. Heat exchange takes place between the $2$ rolls (\SI{60}{\celsius}), the environment (\SI{20}{\celsius}) and the workpiece (\SIrange[range-phrase={ to }]{900}{1100}{\celsius}).

As seen in Figure~\ref{fig:screenshots2DFEM}, re-meshing occurs during the simulation each time the strain change measured from the previous re-meshing exceeds $0.4$. This re-meshing guarantees the reliability and quality of the results. Each 2D FEM simulation takes around \SI{15}{min} to converge~\footnote{2D simulations were performed on the same computing and software platforms as 3D simulations.}.

\subsubsection{3D simplifications}
\label{sec:3dsimplifications}

2D simulations have some drawbacks compared to 3D simulations because the physical problem cannot be reduced to a single cross-sectional representation. The main issues are:
\paragraph{Kinematics of the rolls} The rolls are not rotating, but they are only translating  (closing). The deformation's velocity has a direct impact on the strain rate, which directly influences the material behavior. Therefore the definition of the speed of the closure is of fundamental importance. Compared to 3D simulations, it is clear that the material deforms with speed proportional to the angular velocity of the rolls. Thus, for the 2D case, the rolls are closing with the same vertical speed at which the 3D rolls compress the workpiece.
\paragraph{Stiffness of the workpiece} The 3D workpiece stiffness, during the deformation, moves towards the lateral direction. In the 2D simulation, the cross-section is not fixed in space; therefore, there is no resistance to lateral displacement. For this reason, the profiles of the rolls are chosen, such that friction forces provide enough resistance to lateral movements.
\paragraph{Flowing of the material} Flowing of the material perpendicular to the cross-section is one of the most critical effects for the 3D and the 2D simulations. In the real hot-rolling process, material flows in the cross-section plane as well as perpendicular to such plane, causing a specific reduction of the workpiece cross-sectional area. Usually, this reduction for the first stand reaches values between 20\% and 30\%. We performed 2D simulations with plane stress conditions; it emerged that the reduction of the area for 2D simulations is not congruent with 3D simulations and, thus, not congruent with real profiles. Therefore a model was developed to calibrate the cross-section of 2D simulations with the results of 3D simulations. This model's goal is to predict the inlet diameter of a 3D simulation based on the output of a 2D simulation. Hence, it is possible to run 2D simulations with random inlet diameter and predict the corresponding inlet diameter of the 3D simulation.  
This calibration model can predict the inlet's diameter corresponding to a 3D simulation with an average error of $1.65\%$ and a standard deviation of $1.92\%$.

\subsection{Simulation parameters}

In order to generate random simulations, three inputs are required: temperature, geometry profile of the rolls, and diameter of the workpiece. The temperature of the workpiece is randomly chosen between \SIrange{900}{1100}{\celsius}. The process of generating rolls' geometries and chose the workpiece's diameter is described in Section~\ref{sec:rollgeneration} and Section~\ref{sec:workpiecesdiameter}.

\subsection{Procedural generation of roll geometries}
\label{sec:rollgeneration}

The proposed data-driven approach relies on the availability of large amounts of training data. The 2D simulation strategy detailed in Section~\ref{sec:fesimulationsdesc} allows generating such data.
However, it is required to generate simulations with a variety of deformations. This variation mainly comes from the rolls geometry and the diameter of the workpiece. 
We assumed the stand has a set of horizontal rolls (over and under) and generated it following this procedure:

\begin{figure*}
    \centering
    \includegraphics[width=1.0\linewidth]{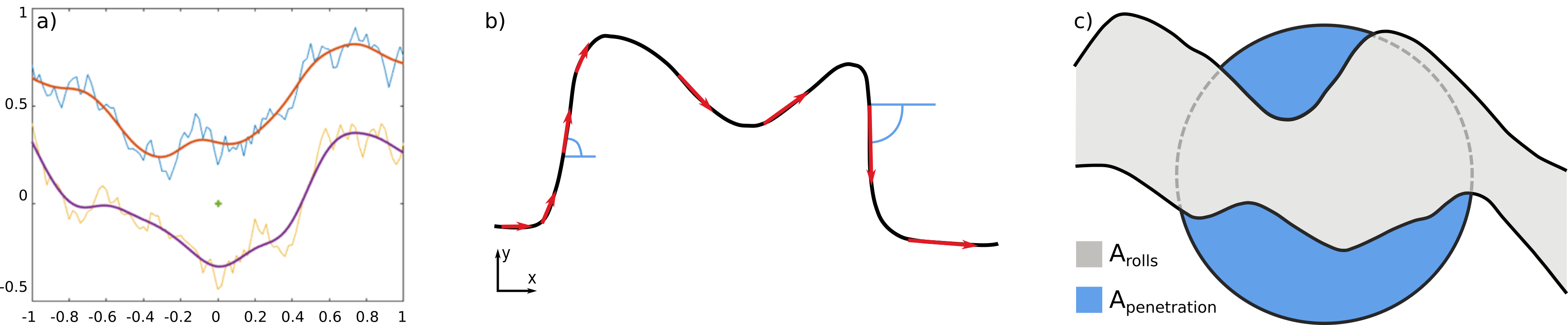}
    \caption{a) Randomly generated lines (yellow for the lower roll and blue line for the upper roll) with their relative spline (red for the upper roll and violet for the lower roll). Illustrations of the rules used in roll generation: b) Undercuts and rule of the draft angle; c) Diameter and penetration area ratio. }
    \label{fig:rollgenerationrules}
\end{figure*}

\begin{enumerate}
\item In a normalized scale, the vector $x$ is created with $101$ values uniformly distributed in $ [-1, 1]$. The vector is ordered with increasing value such that the resulting vector (see Figure~\ref{fig:rollgenerationrules}:b) will never have a component pointing in the negative $x$-direction. This constraint ensures to avoid undercuts in the profile, which in real hot-rolling processes are not possible. Furthermore, the fact that the values are increasing ensures at least a small component pointing in the positive $x$-direction; hence the angle in Figure \ref{fig:rollgenerationrules}:c is always smaller than \SI{90}{\degree} making the process more robust.
\item Random values are assigned to the $y$-vector relative to the over roll (blue line in Figure~\ref{fig:rollgenerationrules}:a.
\item Random values are assigned to the $y$-vector relative to the under roll (yellow line in Figure~\ref{fig:rollgenerationrules}:a). During this vector generation, a minimum distance of $0.4$ is enforced between the over and under profile.
\item The obtained vectors are fitted with splines, depicted in Figure~\ref{fig:rollgenerationrules}:a by violet and red lines.
\item The splines are scaled with the same factor in $x$ and $y$ dimensions to obtain the desired width of the rolls in metric units. The width of the rolls is randomly selected between \SIrange[range-phrase={ to }]{80}{200}{mm}.
\item The vertical distance between the two splines is calculated to ensure a minimum distance of \SI{4}{mm}. If this criterion is not satisfied, the splines are shifted in the vertical direction accordingly.  
\end{enumerate}

Figure~\ref{fig:samplesfem2d} shows some examples of the generated rolls geometries.

\subsection{Workpiece diameter}
\label{sec:workpiecesdiameter}

Once the rolls profiles are generated, the diameter of the input workpiece is randomly chosen among the following: $D = \{20, 24, 28, 30, 34, 38, 42, 46, 48, 50, 52, 54, 56, 58, 60\}$ \SI{}{mm}.  In order to ensure plausible setups, we exclude diameters that do not satisfy the following criteria. First,  the input workpiece area must not be bigger than the area between the rolls (see Figure~\ref{fig:rollgenerationrules}-b): $D \leq 2 \sqrt(\frac{A_{rolls}}{\pi})$; during the deformation of the workpiece, the cross-sectional area becomes smaller (between $20\%$ and $30 \%$ ) due to the material flowing in the out-of-plane direction; this ensures that the area of the workpiece will never fill the area between the rolls. Second, the ratio of the penetration area $A_{penetration}$ (see Figure~\ref{fig:rollgenerationrules}:b) to the area of the workpiece ($\frac{A_{penetration}}{\pi*R^2}$) must range between \SIrange[range-phrase={ to }]{40}{65}{\%}.

\subsection{Validation of 2D simulations}

The validation of our 2D simulation model starts by performing a 2D simulation with a random diameter. Then the inlet diameter corresponding to a 3D simulation is predicted (see Section~\ref{sec:3dsimplifications}). The corresponding 3D simulation is performed using the predicted diameter. Finally, the 2D and the 3D simulation output are compared in terms of the cross-sectional area. Results showed that the cross-sectional area of a 3D simulation is predicted (using the developed 2D model) with an average error of \SI{3.23}{\%}, with the worst-case being \SI{4.91}{\%}.

\section{Experimental setup}

\subsection{Data generation}
\label{sec:datageneration}

In order to generate a sample from a simulation run, a rasterization process is performed to transform the elements in the simulation (workpiece shape, roll geometries, workpiece deformation) into binary images.
Once simulation converges, we consider the curves representing the rolls geometries and the mesh contour representing the deformed workpiece to generate shape rasterizations. Images are generated on the same reference frame, aligned with the initial workpiece center.

\begin{figure}
    \centering
    \includegraphics[width=1.0\linewidth]{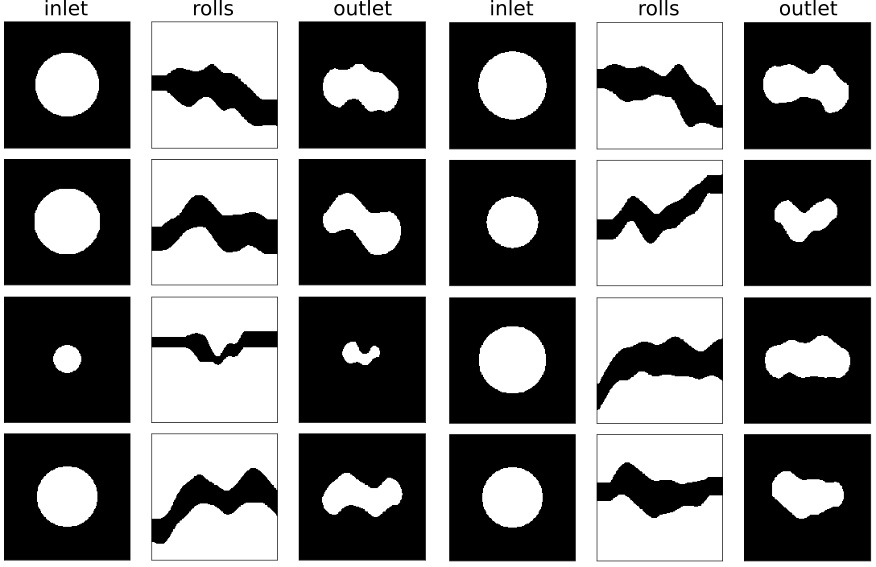}
    \caption{Samples from the 2D FEM simulation. We consider inlets of different sizes and procedurally generated rolls geometries (see Section~\ref{sec:rollgeneration}). The contour of the resulting finite element positions, after simulation converges, represents the deformation.  }
    \label{fig:samplesfem2d}
\end{figure}

Resulting images are of size \SI{200 x 200}{pixels} considering a resolution of roughly \SI{0.5}{mm/pixel}. Each simulation run represents a deformation sample, where the initial shape of the workpiece (circular shape), and the geometry and position of the rolls represent the input; while the deformation (after simulation converges) represents the output. All images are generated considering their origin at the center of the inlet.
Using this sample structure results in a dataset containing only samples for which the inlet is non-deformed, i.e., a circular shape. Focusing on overcoming this limitation, we devised a data augmentation approach where the inlet shape would be a deformed workpiece.
For each simulation run, we analyze the simulation's state at an intermediate frame (see Figure~\ref{fig:screenshots2DFEM}:b). This intermediate deformation is considered the inlet of a new sample while its outlet is the original outlet.
Figure~\ref{fig:samplesfem2d} shows some examples of the samples obtained from the 2D FEM simulation.

\begin{figure}
    \centering
    \includegraphics[width=\linewidth]{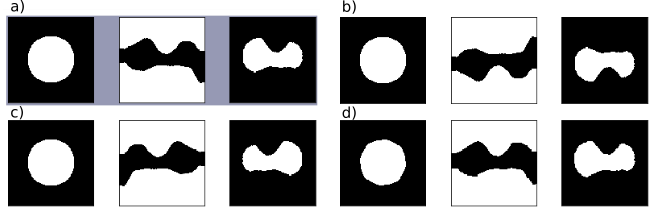}
    \caption{Data augmentation for 2D FEM samples: 
    a) original sample, b) vertical flipping,  c) horizontal flipping, and d) rotation (\SI{-3}{\degree}). }
    \label{fig:dataaugmentation}
\end{figure}

The total 2D FEM dataset is composed of $9400$ samples, that when including the intermediate deformations, increases to $18800$ samples. We randomly split this dataset into training ($14$k samples), validation ($2000$ samples), and evaluation ($2800$).
We implemented an additional data augmentation strategy to increase the training dataset. For each training sample, we flip it vertically or horizontally and slightly rotate it ($4$ times) by an angle sampled from a uniform distribution between \SIrange[range-phrase={ and }]{-3}{3}{\degree}. Hence, the resulting training dataset is composed of around $98$k samples.
Figure~\ref{fig:dataaugmentation} depicts samples of the data augmentation strategy.

\subsection{Deformation model}
\label{sec:deformationestimation}

\begin{figure}
    \centering
    \includegraphics[width=\linewidth]{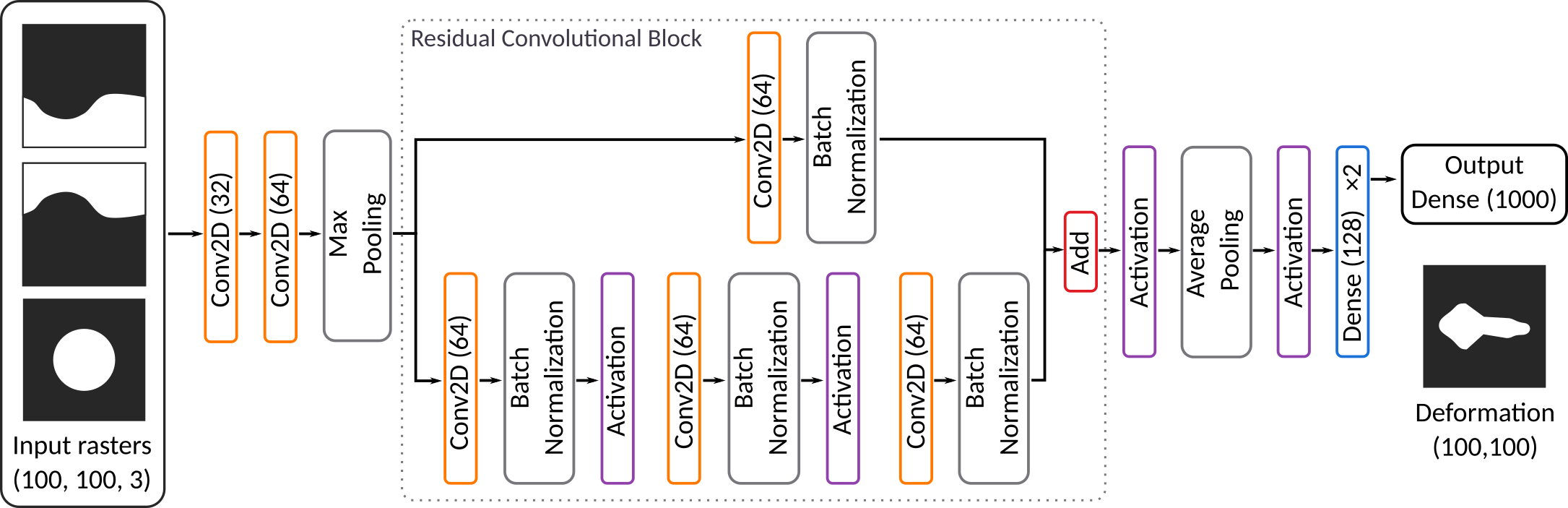}
    \caption{Deformation estimator overview. CNN input consists of geometrical information represented as rasterizations: inlet, over, and under cylinders. The output is the resulting deformation. Our architecture implements a shortcut (Residual Convolutional block)~\cite{he2016}.
    }
    \label{fig:resnetarch}
\end{figure}

The deformation is modeled via a Convolutional Deep Neural Network (CNN), presented in Figure~\ref{fig:resnetarch}. The proposed architecture includes a convolutional residual block that allows the gradient to be directly back-propagated to earlier layers.

Our architecture aims to model the deformation process as a regressor of the outlet image's pixel values. It considers the inlet and the two deforming roll geometries given as a  multi-channel image input. The output represents the deformation at the cross-section of the workpiece and consists of a 2D raster from a dense layer with sigmoid activation.
We use a Mean Square Error metric as the loss function for the training process.

\subsection{Ablated models}

We compare our proposed method against two ablated neural models and two baselines. The ablated neural models consist of two CNN architectures similar to our proposed method but without the Residual Convolutional block. The first architecture (Shallow CNN) contains three hidden convolutional blocks. The second architecture (Deep CNN) consists of eight hidden convolutional blocks.
CNNs integrate features by increasing the number of layers in the architecture, hence the depth in the CNN is of crucial importance to capture relevant features for the task at hand, i.e., deformation. 

\subsection{Baselines}

The baselines employ pixel-wise operations to represent deformations.
The baseline $1$ represents a simplified deformation where the material flow is non-existent; it computes the deformation as the intersection of the gap between the rolls and the inlet shape. The baseline $2$ represents deformations, where the inlet and deformed cross-sectional area remain very similar. The baseline $2$ consists of dilating the inlet shape by a factor $k$, where $k$ represents the diameter in pixels of a circular dilation kernel and then intersecting the dilated input with the gap between the rolls to obtain the output. We repeat for several values of $k \in [2,8]$ and choose the output that best matches the input area. 

\subsection{Performance metric}
We employ the Jaccard similarity index to measure the performance of the proposed estimator against the ground truth. This index measures how similar the target shape is to the estimation. It is defined as the area of their intersection over the area of their union; its values range from $0$ to $1$ when the two shapes perfectly match.

\section{Experimental results}

\subsection{Estimation performance against FEM simulations}
\label{sec:results}

\begin{figure}
    \centering
    \includegraphics[width=\linewidth]{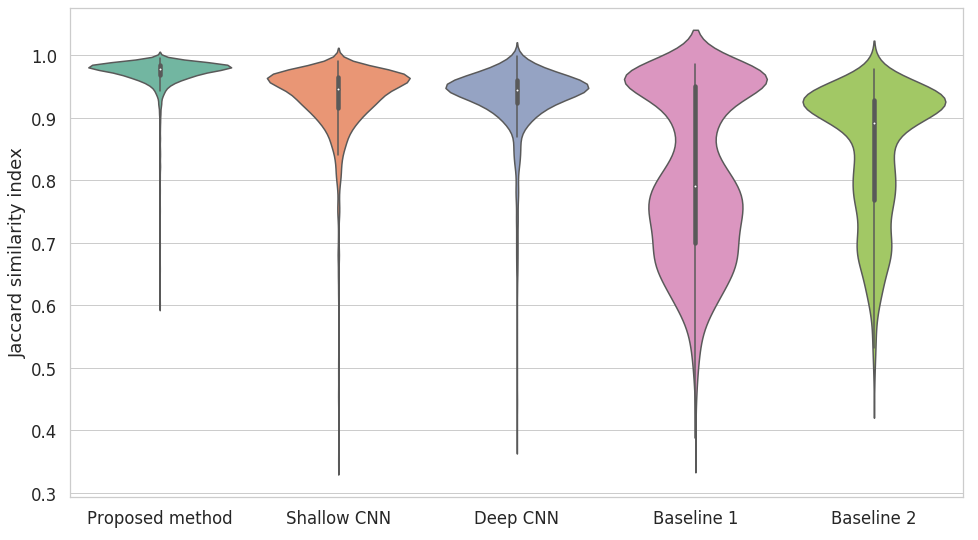}
    \caption{Distributions of Jaccard similarity indexes on the 2D FEM evaluation dataset for the proposed CNN model, Shallow CNN, Deep CNN, and baselines $1$ and $2$. }
    \label{fig:histjaccardsim2dFEM}
\end{figure}

Figure~\ref{fig:histjaccardsim2dFEM} shows the distribution of similarity values when comparing estimates and ground truth values on the evaluation dataset. 
A threshold is applied to the estimated pixel values to obtain a raster of the deformation.

The mean similarity index obtained by the proposed estimator is around $0.972$, which surpasses the mean values from Shallow ($0.932$) and Deep ($0.933$) CNN architectures, and both baselines 1 ($0.805$) and 2($0.843$).
The architecture of the proposed estimator integrates more features needed to capture deformations than that of the Shallow CNN, and its performance does not over-saturate, as is the case of the Deep CNN~\cite{he2016}.
The proposed model can estimate minimal deformations that can be approximated by the intersection operation (baseline 1), and deformations where the material flow perpendicular to the cross-sectional plane is low, i.e., resulting area is similar to the inlet area (baseline 2). The proposed method can also estimate deformations with high material flow variation due to the material being pushed perpendicularly to the cross-sectional plane or toward the rolls gape, which are non-trivial scenarios.

\begin{figure}[h!]
    \centering
    \includegraphics[width=\linewidth]{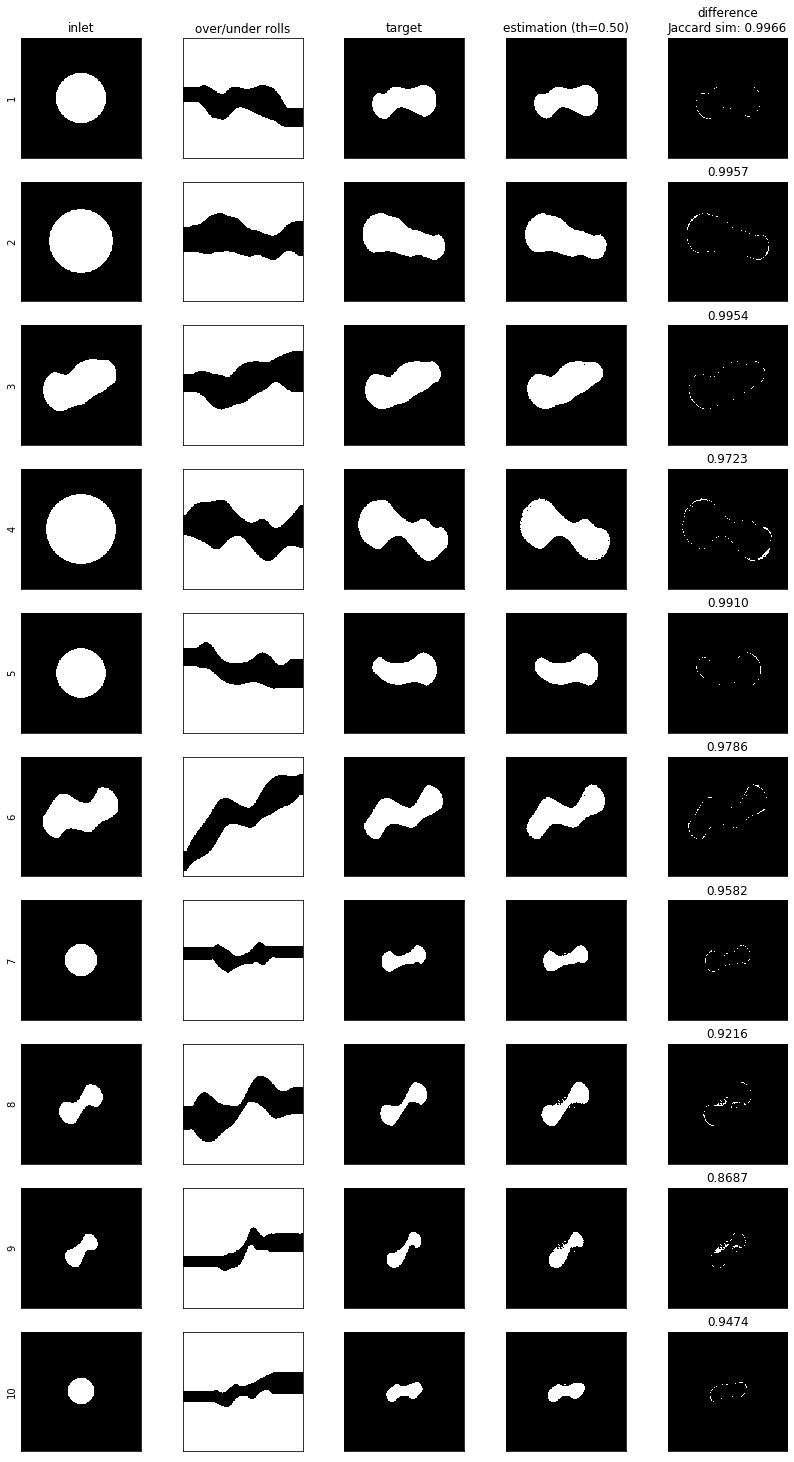}
    \caption{Comparison between the proposed estimator and the ground truth of the 2D FEM simulation evaluation dataset. From left to right: inlet, rolls geometries, target, estimation, pixel difference between target and estimation and, Jaccard similarity index. }
    \label{fig:deformationestimation2DFEM}
\end{figure}

Figure~\ref{fig:deformationestimation2DFEM} presents several examples of estimates obtained with our CNN model and ground-truth from the 2D FEM simulation.
We can observe that estimated deformations are quite similar to the ground truth, in terms of the similarity index. These estimates include cases with deformed (samples 3, 6) and non-deformed inlets (1,2,5).
Challenging regions appear at the sides of the deformation, where part of the material tends to flow when pressed, e.g., samples 4, 6, 10.
There is an error at the contact regions between workpiece and rolls (samples 7, 10). We argue that this error may come from the rasterization process, which down-samples close elements positions to a low-resolution image and appears mainly in small workpieces.

The trained estimator provides modeling of material flow and compression, rendering a finer deformation. However, in some cases, after the threshold application over the estimation's soft values, spurious clusters appear (samples 8, 9). Although this is a direct consequence of the chosen threshold, it can also be the effect of a lack of data for deformations of workpieces with short diameters.
Nevertheless, the deformation's overall shape is correctly estimated and motivates further studies of our proposed method. Computing an estimate takes around \SI{30}{ms}~\footnote{Inference times were obtained on a general-purpose GPU platform  GeForce RTX 2080 Ti.}.

\subsection{Estimation performance against real hot-rolling data}
\label{sec:resultsreal}

\begin{figure}
    \centering
    \includegraphics[width=\linewidth]{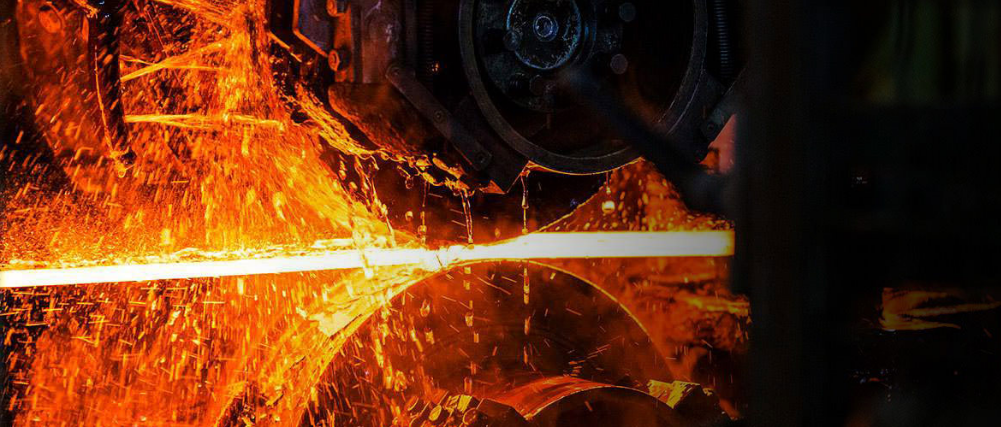}
    \caption{Single-stand hot-rolling with over and under rolls (courtesy of Montanstahl AG).}
    \label{fig:realhotrollingplussamples}
\end{figure}

Hot-rolling involves several steps of tool design and machinery setup. Therefore, real data samples from this process are expensive to create and scarce to obtain. 
In collaboration with Montastahl AG, we gathered a set of 11 single-stand deformation samples from the real process. Such samples comprise input workpieces with a circular shape of diameter between \SIrange[range-phrase={ and }]{38}{51}{mm}. Figure~\ref{fig:realhotrollingplussamples} displays a picture of a real single-stand hot-rolling machinery.

\begin{table}[]
    \ra{1.3}
    \centering
    \begin{tabular}{@{}C{1.1cm}C{2.2cm}C{1.9cm}@{}}
    \toprule
         Sample & Proposed estimator &  Deep CNN\\ \midrule
         1& 0.927  & 0.788  \\
         2& 0.969 & 0.875 \\
         3& 0.948 & 0.824 \\
         4& 0.977 & 0.798 \\
         5& 0.909 & 0.785 \\
         6& 0.921 & 0.847 \\ 
         7& 0.968 & 0.812 \\
         8& 0.887  & 0.851 \\
         9& 0.942 & 0.799 \\
         10& 0.981 & 0.823 \\
         11& 0.975 & 0.810\\
    \bottomrule
    \
    \end{tabular}
    \caption{Jaccard similarity indexes from the proposed estimator and the Deep CNN on real single-stand hot-rolling samples.}
    \label{tab:realdatares}
\end{table}

Table~\ref{tab:realdatares} shows the similarity indexes obtained by the proposed estimator and the Deep CNN compared to the ground truth in the real dataset. 
On these real test samples, we observe good performance; this is very promising for real-world applications. The result is not evident because the model was only trained on procedurally generated samples; however, the rules and constraints used for producing these samples are based on existing hardware tools and material models used in real hot-rolling processes.

\subsection{Preliminary tests on planning of rolling sequences}
\label{sec:planningaplication}

\begin{figure}
    \centering
    \includegraphics[width=\linewidth]{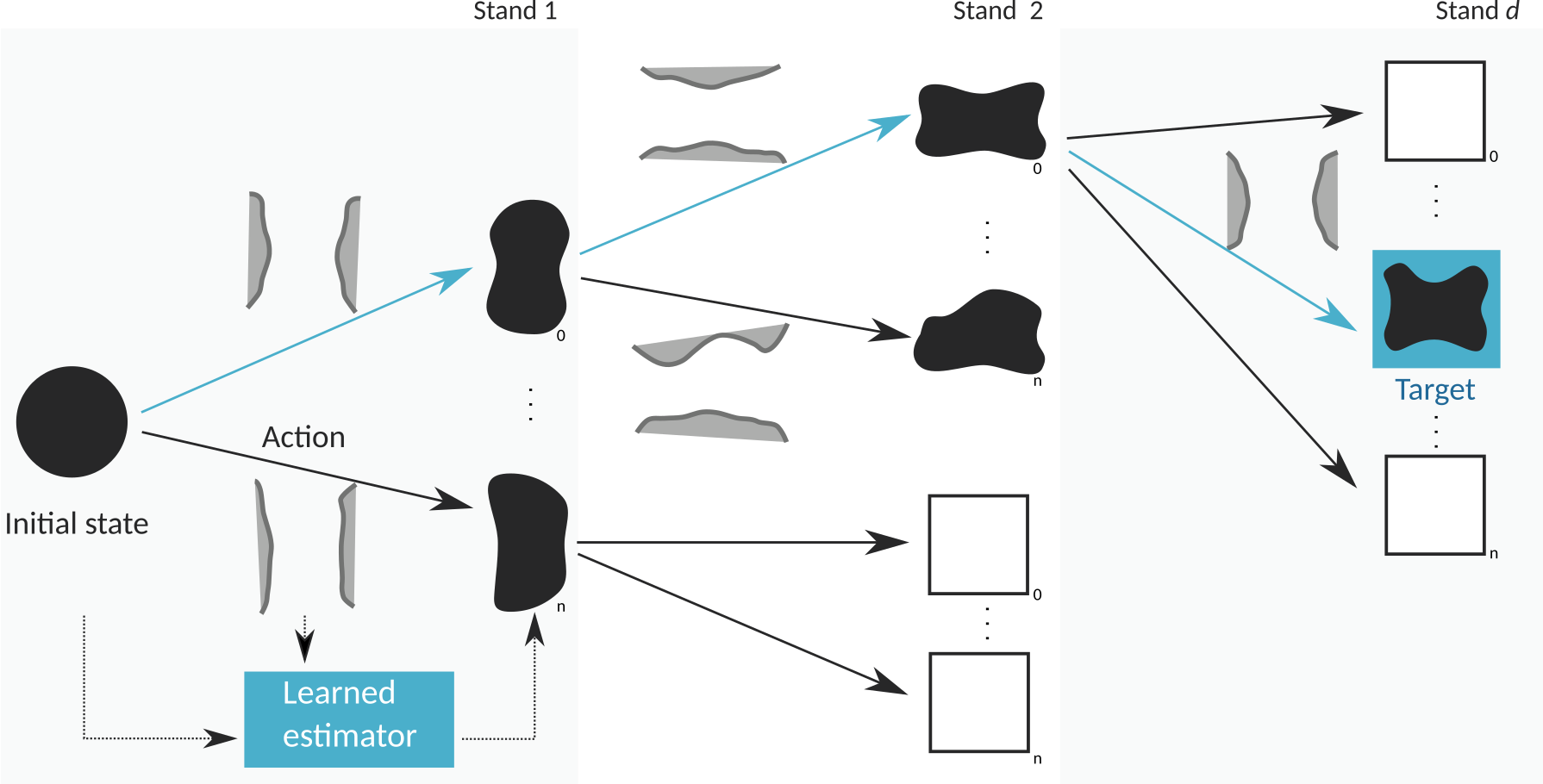}
    \caption{Rolling sequence planning overview. From a query comprised of the initial workpiece geometry and the desired target deformation, we estimate (over several levels) the resulting deformation (outlet) of random roll geometries on the inlet in turn. Then, we select the outlet closer to the desired target and backtrack the sequence of deformations that originated it (blue arrows). Levels in the tree of estimations represent stands in the hot-rolling process.}
    \label{fig:blindplanningstrategy}
\end{figure}

As preliminary work, we tested a simple planning strategy to determine the rolling sequence using our proposed data-drive approach.
Figure~\ref{fig:blindplanningstrategy} depicts an overview of the planning strategy. We start from an initial workpiece (inlet with a circular cross-sectional shape) and the desired target profile. Then we apply $n$ random roll geometries over the inlet to obtain intermediate deformations; done by our proposed estimator (see Section~\ref{sec:deformationestimation}).
We repeat this process for each intermediate deformation until we complete a search tree of depth $d$.
Thus, each depth level represents a stand configuration.
In terms of Jaccard similarity indexes, we compare deformations at each level to the desired target deformation. Then, we select the most similar deformation and backtrack the stand configurations that generate it, i.e., the rolling sequence.

Additionally, as we know the shape of the target profile, we can assume that the final stand configuration (its rolls' geometries) is known. Thus, we also add the final set of rolls geometries for each level in the search tree.
This heuristic steers the blind search towards the desired final configuration and will allow us to explore plans where the desired target is reached before the last level in the search tree.

\begin{figure}
    \centering
    \includegraphics[width=\linewidth]{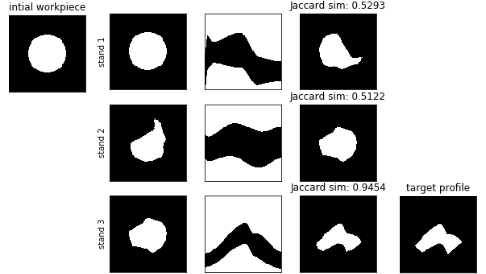}
    \caption{Example of a rolling sequence found by the blind planning approach. From an initial workpiece (top-left) and a target profile(bottom-right), the sequential application of stand configurations (middle rows) results in a deformation similar to the target profile. The outlet of stand 1 was rotated 90 degrees before the stand 2 was applied.}
    \label{fig:exampleresultsplanning}
\end{figure}

Figure~\ref{fig:exampleresultsplanning} shows the result of the planning approach for a given target profile. The initial workpiece has a radius of \SI{32}{mm}.
For each inlet, we explored $100$ random over and under roll geometries and the roll geometries for the final configuration.
The estimator was trained only with over and under roll geometries.
However, by rotating the inlet (\SIlist{90;180;270}{\degree}), we can represent left and right roll configurations. We use this adaptation in the planning approach to explore more complex rolls sequences. At each estimation step, we uniformly choose between rotating or not the inlet in turn.
The search tree has a depth of $3$. 
The closest estimation appears at level $3$ with a coherent rolling sequence. 
Depending on the randomly generated rolls, finding the closest estimation at an earlier level is possible. 
But due to the material displacement in the target profile, a single-stand deformation would not be feasible.
These results show the potential application of our proposed estimator in the designing step of the hot-rolling process. 

Increasing the number of randomly generated rolls per level or the depth of the search tree increases the planning time exponentially. These planning approach experiments were intended as preliminary studies towards a robust planning approach. In previous works, we have explored using an estimator, trained with local information, in a sampling-based global planning framework to identify feasible plans and minimize their costs~\cite{guzzi2020}.

\section{CONCLUSIONS}
\label{sec:conclusions}

We introduced a novel, data-driven approach to estimate metal deformation in a single-stand configuration of the hot-rolling process.

Experimental results show that the proposed CNN-based approach yields deformation estimates similar to a ground truth obtained by state-of-the-art FEM simulations, with a 30000-fold reduction in computational time; the model captures non-trivial physical processes, such as material flow.  The results are also successfully validated against measured deformations from a real-world machine.

This conclusion suggests that learning-based deformation estimators can complement time-consuming FEM simulations in the hot-rolling sequence design process. 
Besides, fast deformation estimations pave the way for automated planning of multi-step rolling sequences that account for process constraints; in this regard, we presented promising preliminary results that motivate our current work in this direction.

As a further extension, we are currently exploring our data-driven approach's capabilities to include additional inputs and outputs relevant to the hot-rolling process, such as temperature, plastic strain on the deformed workpiece, and forces acting on the rolls. This additional information could be used in an automated planning pipeline to estimate the feasibility of a single-stand configuration.

\bibliographystyle{IEEEtran}
\bibliography{references}

\end{document}